\documentclass[12pt]{article}
\usepackage{graphicx}

\begin{document}

\title{Generalized Boltzmann Equation:  Slip-No -Slip Dynamic 
Transition in Flows of Strongly Non-Linear Fluids \\}

\author {Victor Yakhot$^1$, Hudong Chen$^2$, Ilia Staroselsky$^{2}$, \\
John Wanderer$^{1}$ and Raoyang Zhang$^{2}$, 
\\
$^1$ Department of Aerospace and Mechanical Engineering, \\
Boston University, Boston, MA 02215 \\
$^2$ EXA Corporation, 450 Bedford Street, Lexington, MA 02420 \\
}

\maketitle

\begin{abstract}
\noindent The Navier-Stokes equations,  are understood as the result 
of the low-order expansion in powers 
of dimensionless rate of strain $\eta_{ij}=\tau_{0}S_{ij}$,  where  $\tau_{0}$ is the  microscopic 
relaxation time of a close-to- thermodynamic equilibrium fluid. 
In strongly sheared non-equilibrium fluids where $|\eta_{ij}|\geq 1$, the hydrodynamic description breaks down.
According to Bogolubov's  conjecture, strongly non-equlibrium systems are characterized by an  
hierarchy of relaxation times corresponding to  various stages of the relaxation process. 
A "hydro-kinetic"  equation with the relaxation time involving both molecular and 
hydrodynamic components proposed in this paper, reflects  qualitative aspects of  Bogolubov's 
hierarchy.  It is shown that, applied to wall flows, this equation
leads to qualitatively correct results in an extremely wide range of parameter $\eta$-variation. 
Among other features, it  predicts the onset of  slip velocity at the wall as an instability 
of the corresponding hydrodynamic approximation.

\end{abstract}

\vspace{0.3in}

\pagebreak

\section{Introduction}

Strongly sheared fluids, in which the usual Newtonian hydrodynamic description breaks down, 
are commonly encountered in biology, chemical engineering, micro-machinery design [1]-[2]. 
Extensive efforts that largely relied upon physical intuition and qualitative considerations to incorporate 
corrections at the hydrodynamic level of description have been made during  the years 
and have achieved various successes.
However, these attempts generally fail when the non-linearity of a fluid is strong.
Furthermore, theoretical understanding of these highly non-linear systems remains a major challenge.

The Navier-Stokes equations, which  can be derived  from the Boltzmann kinetic
 equation as a 
result of a low -order truncation of an infinite expansion in powers of dimensionless length-scale 
$K_n$ ,  
have  been extremely successful in describing Newtonian fluid flows [3]-[7].
The parameter $K_n$ is defined as a ratio between the so
 called relaxation time $\tau_0$ associated with molecular collisions and the time of the molecular 
convection, namely
\[ K_n = \tau_0 /(L/c_s) = \tau_0 c_s /L \]
where $L$ is a characteristic length scale  a flow inhomogeneity, and $c_s$ is 
the sound speed (i.e., an average speed of the molecules). 
According to kinetic theory, parameter $\lambda=\tau_0 c_s$ represents the so called mean free path
in a system 
and  $\tau_0$ is the time-scale  for the system to relax to its local thermodynamic equilibrium. 
Since the ratio  $L/c_s$ is  a characteristic time-scale  of  deviation from equilibrium due to density  
(concentration) perturbations,  the Knudsen number $K_n$ is 
a measure of  departure  of  a fluid system from thermodynamic equilibrium in  [4]-[6]. 

For the purpose of understanding the effects of non-linearity, it is desirable to re-express $K_n$ in 
terms of the rate of strain (velocity  gradient) in the fluid. Using the estimation for the hydrodynamic (macroscopic) length scale, 
\[ L \sim U/|S| \]
we can also write $K_n$ in an alternative form,
\begin{equation}
K_n = \tau_0 |S| / M_a \equiv \eta_{0} / M_a
\label{kn}
\end{equation}
In the above, $U$ represents the characteristic velocity, and $|S|$ is
 the magnitude of the rate of strain tensor $S_{ij}$ of the flow field ($|S| = \sqrt{2S_{ij}S_{ij}}$). 
$M_a = U/c_s$ is the Mach number. 
The rate of strain tensor is commonly defined as,
\[ S_{ij} = \frac {1} {2} [\frac {\partial u_i} {\partial x_j} 
+ \frac {\partial u_j} {\partial x_i}] \]
The dimensionless rate of strain, $\eta _{0}\equiv \tau_0 |S|$ (in what follows we, instead of $\eta_{0}$, 
 will  use the  parameter $\eta=\tau S$ where 
  $\tau$ is a physically relevant relaxation time) can be viewed as a measure
 of degree of inhomogeneity and shear in the flow. It is equivalent to a 
parameter $\eta \sim W \equiv U\tau_0 /L$ commonly used  in hydrodynamics of polymer solutions.   
Eqn.(\ref{kn}) indicates that $K_n$ can be large if $\eta$ is large, specially for 
low Mach number flows. In many situations $\eta$ is not a small parameter and the Newtonian fluid based 
description breaks down.  For example, capillary flows or blood flows through small
 vessels, rarified gases  and granular flows cannot be quantitatively described as  Newtonian fluids.  A finite $\eta$ can either be a result of a large relaxation time associated with 
the intrinsic fluid property (such as in some polymer solutions), or a result of strong spatial
 variations like  turbulent and micro or nano-flows. This latter effect is particularly important 
at a solid wall where the velocity gradient is often quite significant. It is known that the no-slip 
boundary condition is only valid in the limit of vanishing $K_n$. Experimental data indicate that 
the no-slip condition breaks down when $Kn\geq 10^{-3}$ [8] and the Navier-Stokes description itself 
becomes invalid at $Kn\geq 0.1$. For example, the experimentally observed 
velocity profile in a simple granular 
Couette flow does not resemble the familiar linear variation of velocity predicted from the 
Navier-Stokes equation with no-slip boundary conditions [9].   The wall slip  is an indication of an extremely strong local rate of strain.  

The Navier-Stokes equations can be perceived as a momentum conservation law with a linear 
(Newtonian fluid) stress-strain relation  [3], [5]-[7],  i.e.  the deviatoric part
 of the stress tensor takes on the following form,
\begin{equation}
\sigma_{ij} \equiv \langle {v'_i v'_j} \rangle 
- \frac {1} {3} \langle {{\bf v}'}^2\rangle \delta_{ij}
\approx \nu S_{ij}, \;\;\;\; i,j = x, y, z
\label{nu}
\end{equation}
where coefficient $\nu$ is the kinematic viscosity. 
In the above, ${\bf v}'$ ($\equiv {\bf v} - {\bf u}$) is 
the relative velocity between the velocity of molecules (${\bf v}$) and 
their locally averaged velocity (${\bf u}$) namely the fluid velocity. It is expected that 
${\bf v}'$ and ${\bf u}$ represent, respectively, the fast (kinetic) and the slow (hydrodynamic) 
velocity fields. In an unstrained flow where the rate of strain is equal to zero ($S_{ij} = 0$) , the  
relation (\ref{nu}) is simply interpreted as the first term of an expansion in powers of small 
rate of strain. Therefore, to describe rheological or micro-flows with high rate of strains, such a 
linear approximation must be modified to include non-linear effects. However, this task is
 highly non-trivial if not impossible. 

The hydrodynamic approximations can be
 derived from kinetic equations for the distribution function with the 
intermolecular interactions  accounted  for through the so called collision integral [3], [5]-[7]. 
One can formally write the Bogolubov chain of equations for the distribution function
 and, in addition to the  high powers of the dimensionless rate-of-strain $\eta$, 
 generate an infinite number of equations for  the multi-particle contributions 
to collision integral  [3],[10].
 For a strongly sheared flows  where the  expansion parameter $\eta$ is of order unity, this chain cannot  be truncated. An additional difficulty is that the consistent expansion includes the high-order spatial derivatives ,
 which unfortunately  means that, even if we were able to develop the procedure,  the resulting un-truncated infinite- order hydrodynamic equation requiring  an 
infinite set of initial and boundary conditions 
would be  useless. 

In this paper, based on Bogolubov's concept of the hierarchy of relaxation times, 
 we propose a compact {\it representation} of the infinite- order hydrodynamic formulation in terms of a
simple close-form hybrid (``hydro-kinetic'')  equation. The power of the approach is demonstrated on  a 
classical case of stronly sheared non-Newtonian fluids. 
It is shown that in our ``hydro-kinetic'' approach, the formation of a slip velocity at a solid wall and the  simultaneous  flattening  of velocity distribution in a bulk  is a 
consequence of an ``instability'' of the corresponding  hydrodynamic equation with the no-slip 
boundary conditions. This ``instability'' is the  result of  the non-universal details of 
the flow  such as  the local values of  dimensionless rate of strain $\eta$.  

\section{Basic formulation}

Following the standard Boltzmann kinetic theory, we introduce a density  distribution function 
$f({\bf x}, {\bf v}, t)$ in the so called single particle phase space $({\bf x}, {\bf v})$ ($\equiv {\bf \Gamma}$). A formally exact kinetic equation, 
involving an unspecified collision integral $C$, can be given [11]
\begin{equation}
\partial_t f + v_i \frac {\partial} {\partial x_i} f = C
\label{bolz}
\end{equation}                                                                 
Generally speaking, the detailed expression for $C$ involves multi-particle distribution 
(correlation) functions $f^{(n)}({\bf \Gamma}_1, {\bf \Gamma}_2, \ldots , {\bf \Gamma}_n, t)$ (with $n > 1$). 
This results in the famous Bogolubov chain of infinite number of coupled equations  [3],[10] .
This chain cannot be closed when fluid density is not small.   On the other hand, the collision integral $C$  can be modeled in a relatively simple form in the case of a rarified gas where only the binary collisions are 
important [ 11]
\begin{equation}
C = \int w' (f'f'_1 - f\; f_1)\; d\Gamma_1 d\Gamma' d\Gamma'_1
= \int u_{rel} (f'f'_1 - f\; f_1) d\sigma d^3 p
\label{coll}
\end{equation} 
In this model, $C$ depends only on the single particle distributions.
Here $C \; d\Gamma$ is the rate of change of the number of molecules in the phase volume 
$d\Gamma = d^3 x d^3 v$,  $u_{rel}$ is the relative velocity of colliding molecules and
 $d\sigma$ and ${\bf p}$ are the differential scattering cross-section and momentum of the molecules, respectively. 
The state variable ${\bf \Gamma}$ describes all degrees of freedom 
of a molecule and $w'$ stands for the probability of a transition of 
two molecules initially in states ${\bf \Gamma}$ and ${\bf \Gamma_1}$ 
to states ${\bf \Gamma}'$ and ${\bf \Gamma_1}'$ as a result of collision. 
The kinetic equation (\ref{bolz}) together with the specific collision integral 
(\ref{coll}) forms the celebrated Boltzmann equation. 

It is well known that Boltzmann equation admits an H-theorem in that the system 
monotonically decays (relaxes) to its thermodynamic equilibrium. The corresponding local 
thermodynamic equilibrium distribution 
function, $f^{eq}$, is determined from the solution $C(f^{eq}) = 0$. If deviation from the equilibrium is small, 
we can write 
$f = f^{eq} + \delta f$ and 
\begin{equation}
C = - \frac {f} {\tau_0} + \int u_{rel} f'f'_1 d\sigma d^3 p 
\approx - \frac {f - f^{eq}} {\tau_0}
\label{bgk}
\end{equation}        
with 
\[\frac{1}{\tau_{0}} \approx \int u_{rel} f^{eq} d\sigma d^3 p \]  
Equation (\ref{bgk}) 
is often referred to as the ``BGK'' (mean-field) approximation [12] which 
is a natural reflection of the Boltzmann H-theorem. Furthermore, when perturbation from equilibrium 
is weak, it indicates that the relaxation to equilibrium is realized for each of the distribution 
functions individually having a common relaxation time. At this point, it is important to
 make the following clarification: Even though as shown above that (\ref{bgk}) 
 was deduced from the Boltzmann 
binary collision integral model, its applicability can be argued to go beyond the low density limit. 
Indeed, the model process is consistent to the more general principle of the Second law of thermodynamics: 
A perturbed fluid system always monotonically relax to thermal equilibrium, regardless whether the system has low 
or high density. Furthermore, in accord with Bogolubov [10]
the general process of return-to-local 
equilibrium is true even when the deviation from equilibrium is not small.

In  the classical kinetic theory, the relaxation time $\tau_0$ ($ \approx const$) represents a characteristic time 
of the relaxation process in a weakly non-homogeneous fluid and the  smaller value of $\tau_0$, the  faster the 
process of return to equilibrium.  The formal expansion of kinetic equations in powers of  dimensionless rate of strain (Chapman-Enskog (CE) expansion [3], [6]) developed many years ago, 
leads to hydrodynamic equations for the macroscopic  ("slow") variables.  
When applied to (\ref{bolz}),  the first order trancation of the formally infinite expansion
  gives the Navier-Stokes equations with kinematic viscosity:

\begin{equation} 
\nu_0 = \tau_0 \theta.
\label{nu_0}
\end{equation}

\noindent Development of the expansion based on the full Boltzmann equation is an extremely difficult task. 
On the other hand, the simplified Boltzmann-BGK equation (5) for the single-particle distribution function 
allows extension of the 
Chapman-Enskog expansion to include  higher  powers of dimensionless parameter $\eta$.  
 Considering the higher-order 
 terms of  the Chapman-Enskog expansion, a simple  scalar relaxation term in (5) combined with  advection contribution,  
is expected to generate an
 infinite number of anisotropic contributions as contracted products of $S_{ij}$.  Indeed, by 
expanding equation (5) up to the second order in the Chapman-Enskog series, 
we can explicitly show that the deviatoric part 
of the momentum stress tensor takes on the following form  [13]
\begin{eqnarray}
\sigma_{ij} &=& 
2\tau\theta S_{ij} - 2\tau\theta (\partial_t + {\bf u} \cdot \nabla )(\tau S_{ij})
\nonumber \\
& & - 4 \tau^2 \theta 
[S_{ik} S_{kj} - \frac {1} {d} \delta_{ij} S_{kl} S_{kl} ] 
+ 2 \tau^2 \theta [S_{ik} \Omega_{kj} + S_{jk} \Omega_{ki}]
\label{sed}
\end{eqnarray}
where the vorticity tensor is defined as,
\[ \Omega_{ij} = \frac {1} {2} [\frac {\partial u_i} {\partial x_j} 
- \frac {\partial u_j} {\partial x_i}] \]
Note the first $O(\eta)$  term in (7),  resulting from the first order Chapman-Enskog (CE) expansion,
corresponds to the Navier-Stokes equations for Newtonian fluid,  while 
the non-linear corrections to the Navier-Stokes hydrodynamics are generated in the  next  ($O(\eta^{2})$)  order of the CE expansion.
{\it It is important to further point out that the  memory effects, 
which  appeared in the hydrodynamic approximation (7) 
as a result of the
 second-order CE expansion, are  contained in a simple equation (5) which can be regarded as a generating equation for hydrodynamic models of an arbitrary-order  non-linearity .} \\

The  hydrodynamic approximation (6), (7) has  been  derived from the equation (5) for the single-particle distribution function valid 
for a weakly non-equilibrium fluid (small $\eta$). 
In strongly sheared fluids 
both the assumption of local equilibrium and the low-order trancation of the Bogolubov chain 
(single-particle collisions) fail and the accurate resummation of the series is impossible. 
Indeed, even in the low-density fluids, the  strong shear ($\eta\geq 1$) 
introduces local fluxes facilitating  long-range correlations between  particles,   
invalidating the fluid description in terms of the single-particle distrubution functions. 
To account for these effects, we have to modify kinetic equation (5) in accord with  
some general ideas about the relaxation processes.

\noindent Our goal is to reformulate the  kinetic equation (5),    
 by modifying the relaxation time $\tau$  and come up with the effective kinetic equation 
qualitatively accounting  for the effects of the 
neglected high-order contributions to the Bogolubov chain dominating the dynamics 
far from equilibrium. \\

{\it In his seminal 1946 book [10]
Bogolubov proposed the hierarchy of the time-scales that
 describe relaxation to equilibrium for a system initially far from equilibrium. 
 According to his picture, these  initially   
strong deviations from equilibrium rapidly decrease,  thus  allowing an accurate 
representation of the entire collision integral in terms of the single-particle distribution 
functions. This dramatically simplified representation is sufficient for an accurate 
description  of the later, much slower, process of relaxation to thermodynamic equilibrium. }

To make this plausable assumption operational, we have to represent a Bogolubov 
hierarchy of relaxation times in terms of observable dynamical variables characterizing 
the degree of  deviation from equilibrium. 
Since the most natural parameter governing the dynamics far from equilibrium is 
the rate of strain 
$|S_{ij}|$,  we propose that both the close-to-equilibrium relaxation time 
 $\tau_{0}$ and inverse rate-of-strain $1/|S_{ij}|$ define the Bogolubov hierarchy of the relaxation times. 
The simplest Galileo invariant relaxation model that is compatible with the above physical considerations is:
\begin{equation}
\frac {1} {\tau} = \sqrt{\frac {1} {\tau_0^2} + \gamma |S^{ne}|^2}
\equiv \frac {1} {\tau_0} \sqrt{1 + \gamma \eta_{ne}^2}
\label{tau}
\end{equation}

\noindent where we define in accord with (2), 
$S^{ne}_{ij}\approx \frac{\sigma_{ij}}{\nu_{0}}$ and $\eta_{ne}=\tau_{0}|S^{ne}|$ ($i\neq j$). 
In what follows, wherever it cannot lead to misunderstandings, 
 we will often omitt the suffix $ne$. The new collision integral (i.e., (\ref{bgk}) with  
$\tau_0$ replaced by $\tau$ in 
(\ref{tau})) 
now describes a 
relaxation process that involves a rate of strain-dependent 
relaxation time.

 {\it The proposed hydro-kinetic model ((\ref{bolz})-(\ref{tau})) is chosen to reflect 
some of the principle elements in the Bogolubov hierarchy. That is, far from
 equilibrium where the rate of strain is large, the essential time-scale $\tau$ is dominated by 
$|S|^{-1}$ which corresponds to a rapid first stage of the relaxation process. Later on,
 when the rate of strain becomes small, the relaxation process is governed by a 
close-to-equilibrium relaxation time $\tau_0$, as used in the conventional BGK equation (5) leading 
to the Navier-Stokes formulation.}
It is clear from relation (7) that even though, by restricting our model 
to the scalar relaxation times in (\ref{tau}),  the anisotropic contributions to the stress do appear 
in the hydrodynamic description 
which is a result of the Chapman- Enskog expansion. 
Since the rate of strain is a property of the flow, the model (8) which includes both 
molecular and hydrodynamic features can be called a hybrid hydro-kinetic approach to non-linear 
fluids. 
 It is shown below that even with such a specific form in (\ref{tau}), 
the hydro-kinetic equation
is capable of producing some quite non-trivial but physically sensible results at the hydrodynamic level.

To sumamrise the main points of this section: 
it is analytically impossible and practically useless to attempt to 
describe 
the strongly non-linear (non-equilibrium) flow physics at the hydrodynamic level. 
As indicated above, not only 
the resulting ``differential equation'' does not have a finite form, it also requires an infinite number 
of boundary conditions. The reason for this fundamental difficulty is that the expansion in powers of 
 parameter $\eta$ becomes invalid when $\eta$  is not small. Thus, to deal with strongly sheared and
 time-dependent non-linear flows, it is desirable to use the simple ``hydro-kinetic'' description 
(\ref{bolz})-(\ref{tau}). Clearly,  this hybrid representation has a finite form, while it formally
 contains all the terms in the infinite expansion for the hydrodynamic level. As argued above, the 
hydro-kinetic formulation is applicable to both large and small $\eta$, corresponding to large and
 small deviations from equilibrium.

\section{Wall flows of non-Newtonian fluids}

To illustrate the benefits of the hybrid representation of (\ref{bolz})-(\ref{tau}), let 
us first consider a laminar unidirectional flow in a channel between two plates separated by distance 
$2H$ and driven by a constant gravity force $g$. In a steady state, the Navier-Stokes equation for
 the channel flow can be derived from (5), (8) in the lowest order of the expansion in powers of dimensionless 
rate of strain $\eta=\tau|S_{ij}|$. Repeating the procedure leading to (7) and restricting 
the expansion by the first term gives the Navier-Stokes equation 
having a "renormalized" viscosity corresponding to (\ref{tau}) ,
\begin{equation}
\partial_{t}u- g = \partial_y {\tilde \nu} \partial_{y} u
\label{ns}
\end{equation}
where
\begin{equation}
{\tilde \nu} = \tau\theta = \nu_0 /\sqrt{1 + \gamma \tau^2_0 |S|^2}
\label{nun}
\end{equation}
for the special unidirectional situation, $|S| = |\partial u/\partial y |$. 
In the above, we have chosen $u$ to be the streamwise velocity component, 
while coordinates $x$ and $y$ are along the streamwise and normal 
directions of the channel, respectively. 

One important point must be mentioned: the expansion in powers of $\eta_{0}=\tau_{0}S$ corresponds to the classic CE expansion. The equation (9) has been derived by expanding in powers of
$\eta=\tau |S| =\eta_{0}/\sqrt{1+\gamma \eta_{0}^{2}}$, which means that even the low-order perturbation theory in powers of the "dressed " parameter $\eta$ corresponds to an infinite series in powers of  "bare" parameter $\eta_{0}$. Thus, it is extremely interesting to assess the accuracy of the derived hydrodynamic approximation (9)-(10).

Subject to no-slip boundary conditions, the exact 
steady-state analytic solution for (\ref{ns}) is given by:
\begin{equation}
u(y) = \frac {gH^2} {\nu_0\beta} 
(\sqrt{1 - \beta (\frac {y} {H} )^2} - \sqrt{1 - \beta} )
\label{soln}
\end{equation}
where $\beta \equiv \gamma (\tau_0 g H/\nu_0)^2$ is a dimensionless parameter which can 
be either positive or negative, for $\gamma$ can in principle have either positive or negative
 signs. One can immediately see that this particular flow solution breaks down for $\beta > 1$. 
A direct indication of this is that the steady state Navier-Stokes equation (\ref{ns}) is no longer 
valid to describe such a flow, and we have to return back to the full hydro-kinetic representation 
(\ref{bolz})-(\ref{tau}). It is interesting that the equation (9)-(10) allows an unsteady singular solution:

\begin{equation}
u=gt+u_{0}\phi(y)
\end{equation}
\noindent where $\phi(y)=1$ for $|y|<H$ and $\phi(y)=0$ on the walls where $|y|=H$.
The transition  between the two (no-slip (11) and slip (12)) solutions will be demonstrated below.

In the rest of the paper, we present solution of the full hydro-kinetic 
system for the channel flow for the entire range of parameter variation of $\gamma$ with initial velocity profile $u(y,t=0)=0$.  For this purpose,
 equations (\ref{bolz})-(\ref{tau}) have been numerically solved using a Lattice Boltzmann (LB) 
algorithm [14]. 
On each time step 
the relaxation time in (8) was calculated 
with  the non-equilibrium rate of strain $S^{ne}_{ij}$, defined as:
\begin{equation}
S^{ne}_{ij}=\frac{1}{\nu_{0}}\int dv_{i}dv_{j}(v_{i}-u_{i})(v_{j}-u_{j})(f-f^{eq})
\end{equation}

\noindent with $i\neq j$ and ${\bf u}$ standing for the local value of the 
mean velocity. It is clear from this definition
that in thermodynamic  equilibrium, the rate-of-strain 
$S_{ij}^{eq}=0$ and, according to (7),  not far fom equilibrium  $S_{ij}\approx S^{ne}_{ij}$.
The computationally effective and widely tested 
 ``bounce-back'' collision process giving, on the hydrodynamic level of description, 
rise to the no-slip boundary condition in the $K_n\rightarrow 0$ limit, was imposed on a solid wall. 
According to the "bounce back"  algorithm, the  momentum of the "molecule"
colliding with a  solid surface changes according to the rule: 
${\bf p}\Rightarrow -{\bf p}$.

When parameter $\beta=0$ , the familiar steady state parabolic solution 
$u(y)=\frac{gH^{2}}{\nu_{0}}(1-(\frac{y}{H}^{2})$ was readily derived.
Figs.1 present the analytical (i.e., (\ref{soln})) and the numerical (i.e.,
 (\ref{bolz})-(\ref{tau})) solutions of the velocity profiles in the plane channel flow
 for, respectively $\beta = 0.4$. 
As we can see, for this 
 value, the difference between the solutions of (\ref{soln}) and simulations of the full 
hydro-kinetic model ((\ref{bolz})-(\ref{tau})) is negligible. 
This means that in this regime, 
the lowest order trancation of the Chapman-Enskog expansion in powers of the "dressed" parameter $\eta$ (7) is extremely accurate. 
The same conclusion was shown to hold for all $\beta<0.5$. 
\begin{figure}[h]
  \center
\includegraphics[height=8cm]{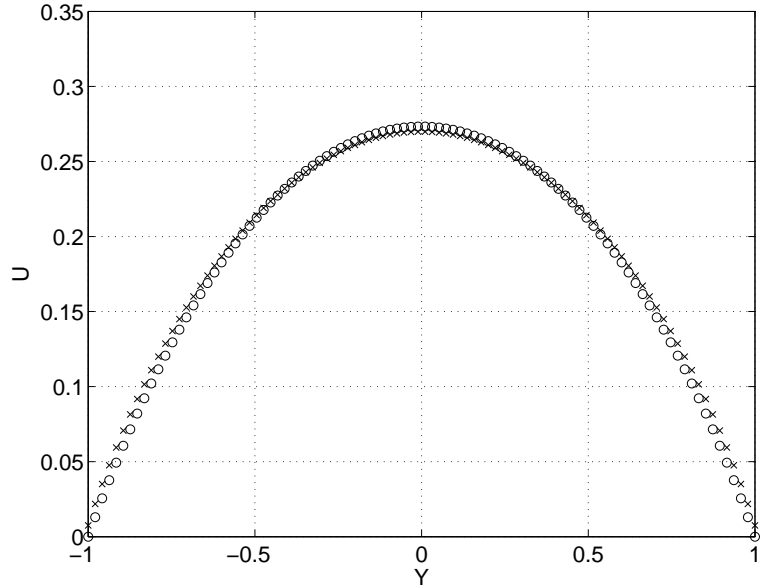}
  \caption{Steady state velocity profile: comparison of hydrodynamic solution  Eq. 11 ($\times$)  with the LBM simulation $\circ$.  $\beta=0.4$. }
  \label{fig:stable}
\end{figure}

The numerical results for $\beta\geq 0.51$ revealed an interesting instability theoretically predicted , for $\beta\geq 1$. The results of very accurate numerical simulation (960 cells across the channel) are presented on Figs. 2-3. 
\begin{figure}[h]
  \center
  \includegraphics[height=8cm]{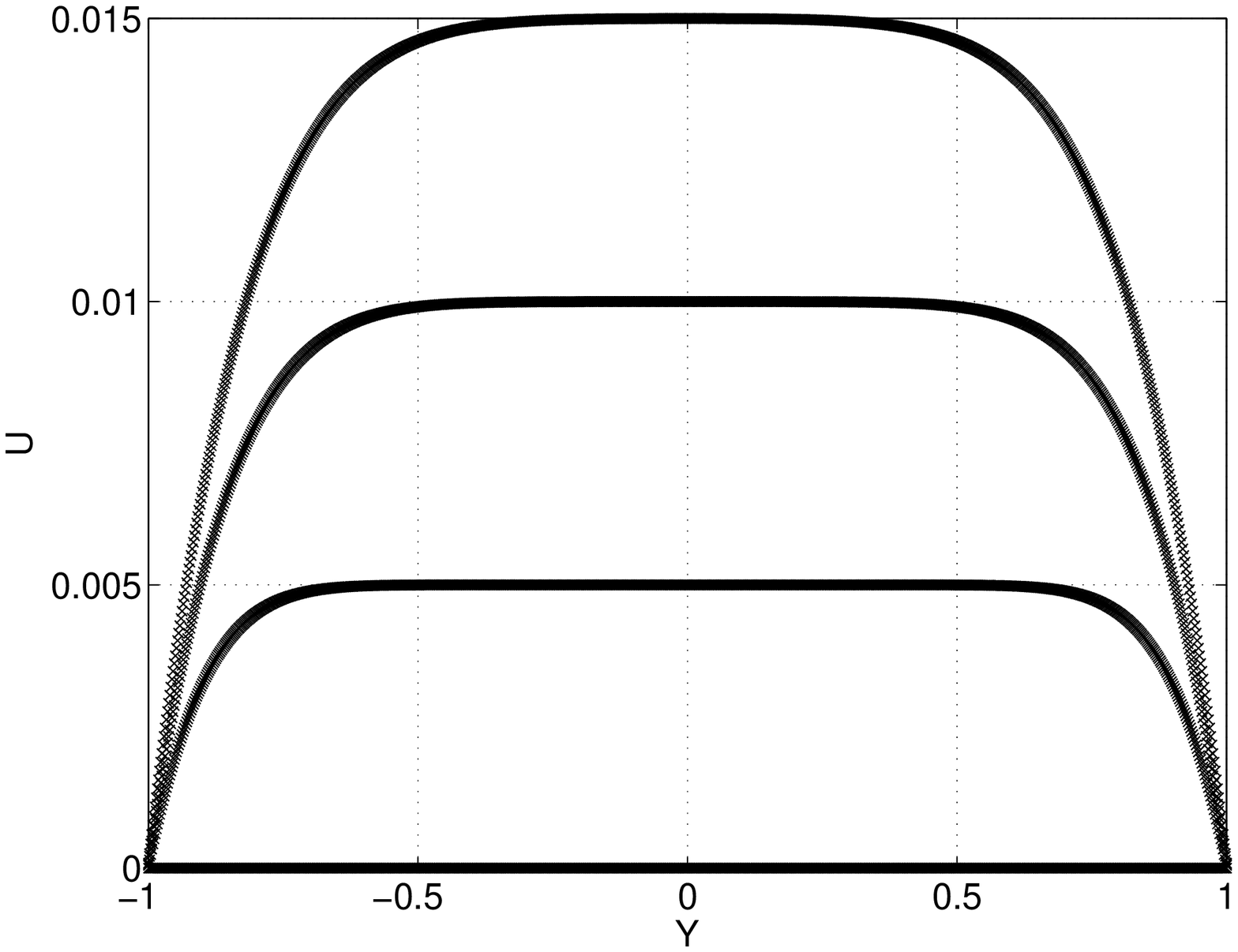}
  \caption{Short-time evolution of unstable velocity profile $\beta=0.51$. The values of dimensionless time in arbitrary units: from bottom to top: T=0;~2;~4;~6. }
  \label{fig:unstable}
\end{figure}

We can notice
 qualitatively new features not captured by the steady-state hydrodynamic approximation: 
initially,  formation of aly narrow wall-boundary layer,  accompanied by 
 a strong flattening of the velocity profile in the bulk can be observed.

\begin{figure}[h]
  \center
 \includegraphics[height=8cm]{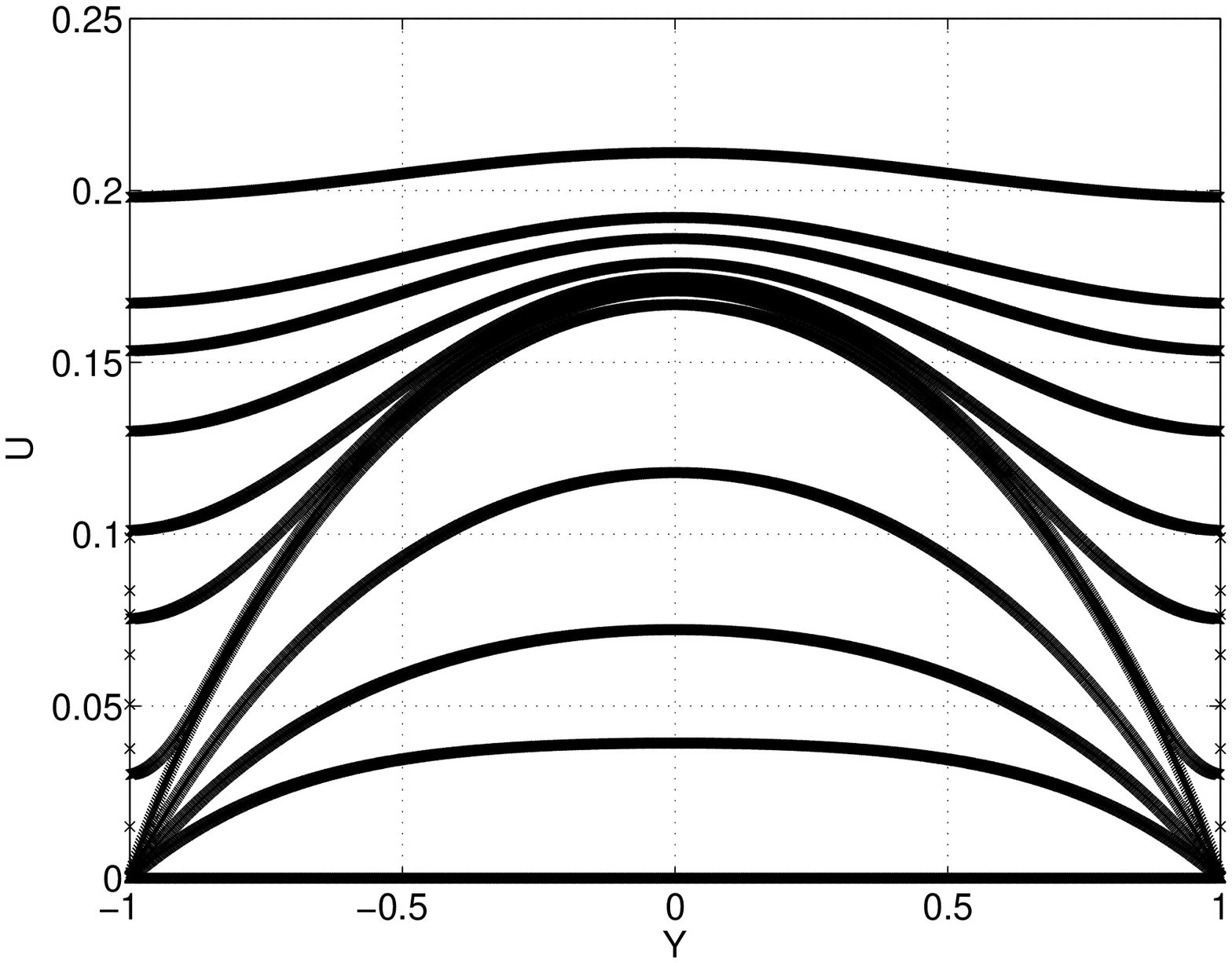}
  \caption{Long time velocity profile evolution $\beta=0.51$. 
  The values of dimensionless time in arbitrary units: from bottom  to top: T=0;~8;~16;32;.........80.}
  \label{fig:unstable_evolution}
\end{figure}

Later in evolution, the boundary layer becomes unstable with formation of the slip velocity at the wall.
The flow accelerates,  eventually becoming a free -falling plug  flow,  predicted by equation (12). 
 All these phenomena have been experimentally observed in the flows of  rarified gases and granular materials.
 
A clarification of the set up is in order: The simulations
 were performed on an effectively infinitely  long (periodic boundary conditions along the streamwise direction)
 channel, and the flow was  driven by an externally imposed constant gravity. 
This set up differs quite substantially from a pressure-gradient-driven flow of a nonlinear fluid where a
 steady state can be achieved by formation of a non-constant ($x$-dependent) streamwise pressure gradient.
 Unlike pressure,  gravity is not a dynamical variable and hence the flow lacks the mechanism for establishing a
 force balance needed to achieve a steady state. This can be associated with the experimentally observed inability of  the gravity-driven granular flows in ducts
 to reach steady velocity profiles  [14].
In accord with this theory, the steady velocity profile similar to those
shown on Figs. 2-3 can also be observed in a gravity- driven finite-length-pipe or channel flows . In this case we expect the velocity distribution to vary with the length of the pipe/channel.
\newpage

\section{Conclusion}
  
It has recently  been  shown that the Lattice Boltmann ( BGK ) equation with the  effective strain-dependent relaxation time can be used  for accurate description of  high Reynolds number turbulent flows in complex geometries [16].  In this work,  this concept has been generalized to flows of strongly non-linear fluids.
Although the simple relaxation model (\ref{tau}) was proposed 
here  on a qualitative basis,
 it has shown to be capable of producing
non-trivial predictions for flows involving strong non-linearity.  To the best of our knowledge,  this hybrid ("hydro-kinetic") model 
is the first attempt of  incorporating  the principle elements of the Bogolubov conjecture about infinite hierarchy of relaxation times.  The most interesting result of application of this model is the 
appearance of the slip velocity on the wall as a result of dynamic transition driven by increasing rate of strain.  Since this transition depends on the wall geometry,  it cannot be universal.  Thus, to predict  this extremely important effect, the model (3)-(8) does not require empirical, externally imposed  boundary conditions. The classical incompressible hydrodynamics relies upon one externally determined parameter, the viscosity coefficient which can be obtained either theoretically (sometimes) or  from experimental data.    The hydro-kinetic approach proposed in this paper needs a single additional  parameter $\gamma$ describing  physical properties of a strongly non -linear fluid , which 
can readily be established from  a   low Reynolds number  
flow in a capillar   by comparing the measured velocity profile with the theoretical prediction (11).
Further applications of the model (3)-(8)  to the   separated highly non-linear flows,  will show how far one can reach using this simple approach.

{\bf Acknowledgements}. 
One of the authors (VY) has greatly benefitted from stimulating discussions with R.
 Dorfman, I. Goldhirsch, K.R. Sreenivasan, W. Lossert, D. Levermore, A.Polyakov.

\end{document}